\documentstyle[12pt]{article}
\begin{document}
\title{On instability of hadronic string \\ with heavy quarks at its ends}
\author{G. Lambiase\thanks{Permanent address:
Dipartimento di Fisica Teorica
e S. M. S. A., Universit\'a di Salerno,
84081 Baronissi (SA), Italia},~~~V.V. Nesterenko \\
{\small \it Bogoliubov Laboratory of Theoretical Physics}\\
 {\small \it Joint Institute for Nuclear Research, Dubna, 141980, Russia}}
\date{}
\maketitle
\begin{abstract}
The quark mass dependence of the energy spectrum in the Nambu--Goto 
string with point--like masses (quarks) at its ends is analyzed.  To 
this end, linearized equations of motion and boundary conditions in 
this model are considered. It is shown that for sufficiently large 
quark masses, the first excited state in string spectrum may be 
arbitrary close to the ground state. Obviously this points to 
infrared instability in the system under consideration. Possible 
modifications of string model which could remove this drawback are 
discussed.
\end{abstract}
\newpage

\vspace*{2cm}

\section{Introduction}

Investigation of the quark dynamics inside hadrons remains as 
before the most
important task of theoretical and experimental studies in hadronic 
physics. In
view of asymptotic freedom in QCD, this problem at small distances
can be treated in the framework of perturbative calculations.
However at distances like hadron size $R_{h}$ and larger 
nonperturbative effects become important. 
Therefore in this region lattice simulations and 
string models are used for revealing the interquark interaction 
(for a review and further references, see, e.g. ~\cite{FLEN},~\cite{BORN}).

Usually one assumes that at distances $\sim R_h$  the most important 
are such configurations of gluonic field when this field is 
concentrated along the line connecting quarks (the flux tube between 
quarks). Such infinitely thin flux tube is simulated by relativistic 
string. In the most simple case, the dynamics of this system is 
determined by the Nambu--Goto string action ~\cite{BARB}. However
other string models were also proposed to describe quark interaction, 
for example Polyakov--Kleinert rigid string ~\cite{POLY},~\cite{KLEI}.
Search for new string models
is stimulated by comparing the calculations in QCD with those in the 
framework of known string models ~\cite{POLC}.

A large body of literature is devoted to the calculation of the 
interquark potential in string models (see, for example, 
~\cite{ALVA}--\cite{ELIZ} and references therein). The central
point in these calculations is a determination of the string energy 
as a function of the distance between quarks. When deconfinement 
temperature (or critical temperature) is investigated then the 
dependence of the effective string tension on temperature should be 
calculated. Practically in all these investigations only the static 
interquark potential has been considered\footnote{First attempt to 
calculate the quark mass dependence of the string potential was made 
in Ref.~\cite{UNPU}. The authors are grateful to Prof. H. Kleinert 
for providing them with this paper.}.
It means that the ends of the string connecting quarks are fixed. 
On the one hand, this assumption essentially simplifies the problem 
because the string frequencies in this case are integer. On the other 
hand, results obtained in this way are applied, strictly speaking, 
only to the infinitely heavy quarks which cannot move. However in 
this case the notion of the interquark potential practically loses 
its usefulness.

As a matter of fact static string potential is treated, after its 
derivation, as a usual potential describing the interaction between 
quarks with finite masses. Here it is implicitly assumed that the 
interquark potential is a smooth function of quark mass $m$ when 
$m\to\infty$. Probably it is right. Calculation of the interquark 
potential generated by string with massive ends
carried out in our recent paper ~\cite{LANE} testifies to this 
conclusion. Nevertheless we would like to draw attention to one 
peculiarity in energy spectrum of the relativistic string with heavy 
quarks at its ends which is, to our view, very important. The point 
is that the position of the first excited level in this system depends 
on the quark mass $m$ and when $m\to\infty$ this level may be arbitrary 
close to the ground state. Obviously this testifies to the infrared 
instability of the string connecting quarks with sufficiently large 
masses. It is this problem that will be discussed in our short 
comunication. The layout of the paper is the following. In Section 2 
the basic facts from the classical and quantum theories of the 
Nambu--Goto string with massive ends are given. Then linearization 
of equations of motion and boundary conditions is considered. It 
should be noted here that it is these linearized equations that are 
used as a starting point in  all the calculations of the interquark 
potential in string models both in perturbation theory
and by making use of the functional integration (variational 
estimations). Therefore the spectrum of eigenfrequencies in this 
system is of great importance. In Section 3 the quark mass dependence 
of the string spectrum is investigated. It is shown that the energy 
of the first excited level, $E_1$, may be arbitrary close to the ground 
state with energy $E_0$ when quark mass tends to infinity. This is a 
direct indication about the infrared instability in the system under
consideration. In Conclusion (Section 4) the physical implications of 
this result and possible modifications of the string model for
removing this drawback are shortly discussed.

\section{Equations of motion and boundary conditions in linear 
         approximation}
\setcounter{equation}{0}

The Nambu--Goto string with point--like masses attached to its ends 
is described by action~\cite{BARB}
\begin{equation}
S\,=\,-M_0^2\,\int\!\!\!\!\!\int\limits_{\Sigma\hspace*{0.25cm}}^{}
d\Sigma\,-\,
\sum_{a=1}^{2}\,m_a\!\int\limits_{C_a}^{}\!\!ds_a\,{,}
\end{equation}
where $d\Sigma$ is infinitesimal area of the string world surface,
$C_a$, $a\,=\,1,2$, are the world trajectories of the string massive 
ends, $m_a, a\,=\,1,2,$ are the quark masses ($\hbar\,=\,c\,=\,1$).
$M_0^2$ is the string tension. In hadronic physics one usually sets 
$M_0\,\sim\, 1$~GeV.

The Gauss parametrization of the string world surface embedded into
$D$--dimensional space--time with signature $(+,-,\dots,-)$

$$ x^{\mu}(t,r)\,=\,(t,r;x^1(t,r),\dots,x^{D-2}(t,r))\,= $$
\begin{equation}
=\,(t,r;{\bf u}(t,r))\,{,}
\end{equation}
permits us to write the induced metric on this  surface,
$g_{ij}\,=\,\partial_i x^{\mu}\partial_j x_{\mu}$, in the following 
way
\begin{equation}
g_{ij}\,=\,\delta_{ij}\,-\,{\bf{u}}_i{\bf {u}}_j\,{,}
\quad i,j\,=\,0,1\,{.}
\end{equation}
Here ${\bf u}_0\,=\,\partial_0{\bf u}\,=\,
\partial{\bf u}/\partial t\,=\,\dot{\bf u},\;\;
{\bf u}_1\,=\,\partial_1{\bf u}\,=\,\partial{\bf u}/\partial r\,
=\,{\bf u}^{'}$ and ${\bf {u}}{\bf {u}}\,
=\,\sum_{j=1}^{D-2}\,u^j u^j$. From Eq.~(2.3) it follows that in the
quadratic approximation the determinant $g$ of the induced metric is 
given by
\begin{equation}
g\,=\,\det(g_{ij})\,\simeq\,1\,-\,{\bf {u}}_i^2\,{.}
\end{equation}
In this approximation, the infinitesimal area $d\Sigma$ can be written 
as $d\Sigma\,=\,\sqrt{g}\,dt\,dr\,\simeq\,[1\,-\,{\bf {u}}_i^2/2]
\,dt\,dr$, while the line elements $ds_a,\, a=1,2,$ take the form 
$ds_a\,=\,[1\,-\,\dot{\bf {u}}^2(t,r_a)/2]\,dt$. Finally the action 
(2.1) becomes
\begin{equation}
S\,\simeq\,\frac{M_0^2}{2}\,\int\limits_{t_1}^{t_2}\!dt
\int\limits_{0}^{R}\!dr\,\left[\dot{\bf {u}}^2(t,r)\,-\,
{\bf {u}}^{'2}(t,r)\right]\,+\,\sum_{a=1}^{2}\,
\frac{m_a}{2}\,\int\limits_{t_1}^{t_2}\!dt\,
\dot{\bf {u}}^2(t,r_a)\,{,}
\end{equation}
$$r_1\,=\,0,\quad r_2\,=\,R\,{.}$$

Equations of motion and boundary conditions for the system described 
by (2.5) can be immediately deduced
\begin{equation}
\Box {\bf u}(t,r)\,=\,0 \,{,}
\end{equation}
\begin{equation}
m\,\ddot{\bf u}\,=\,M_0^2\,{\bf {u}}^{'}, \qquad r\,=\,0\,{,}
\end{equation}
\begin{equation}
m\,\ddot{\bf {u}}\,=\,-\,M_0^2\,{\bf {u}}^{'}, \qquad r\,=\,R\,{,}
\end{equation}
where $\Box\,=\,\partial^2/\partial t^2\,-\,\partial^2/\partial r^2$ 
and we put for simplicity $m_1\,=\,m_2\,=\,m$.
The general solution to Eq.~(2.6)--(2.8) is given by
\begin{equation}
u^j(t,r)\,=\,i\,\frac{\sqrt{2}}{M_0}\,\sum_{n\not= 0}^{}\,
\exp{\left[-iM_0\omega_n t\right]}\,\frac{\alpha^j_n}{\omega_n}\,
u_n(r),\quad j=1,\dots, D-2\,{.}
\end{equation}
Amplitudes $\alpha^j_n$ satisfy the usual rule of complex
conjugation $\alpha^*_n\,=\,\alpha_{-n}$. The eigenfunctions $u_n(r)$ 
in (2.9) are defined by
\begin{equation}
u_n(r)\,=\,N_n\,[\cos{(\omega_nM_0r)}\,-\,
\mu\omega_n\sin{(\omega_nM_0r)}]\,{.}
\end{equation}
$N_n$'s are normalization constants and $\mu$ is the dimensionless 
parameter
\begin{equation}
\mu\,=\,\frac{m}{M_0}\,{.}
\end{equation}
The eigenfrequencies $\omega_n$ are the roots of the transcendental
equation\footnote{This equation is well known in mathematical physics.
Besides the
problem about linear vibrations of the string with point--like masses 
at its ends, this equation determines eigenfrequencies of the 
torsional vibrations of shaft with massive discs at its ends 
~\cite{TIMO} and eigenvalues in the boundary value problem about heat 
flow along rod under special conditions at its ends~\cite{KOSH}\,{.}}
\begin{equation}
\tan(\omega_nM_0R)\,=\,\frac{2\mu\omega_n}{\mu^2\omega_n^2\,-\,1}\,{.}
\end{equation}
On the $\omega$-axis these roots are placed symmetrically around zero. 
Hence they can be numbered in the following way $\omega_{-n}\,=\,-
\omega_n,\,n\,=\,1,2,\dots.$ Therefore it will be sufficient to 
consider only the positive roots. The Hamiltonian operator of the 
system under consideration in terms of creation and annihilation 
operator, $a^{+}$ and $a$ respectively, reads
\begin{equation}
H\,=\,M_0\sum_{n=1}^{\infty}\,\sum_{j=1}^{D-2}\,
\omega_n\,a^{j+}_na^{j}_n\,
+\,\frac{D-2}{2}\,M_0\,\sum_{n=1}^{\infty}\,\omega_n\,{,}
\end{equation}
where
$$[a^i_n,\,a^{j+}_m]\,=\,\delta^{ij}\,\delta_{nm},\quad 
n,m\,=1,2,\dots \,{.}$$
The last term in Eq.~(2.13) is the Casimir 
energy ~\cite{MOST}, ~\cite{PLUN}.

\section{Quark mass dependence of the string spectrum}
\setcounter{equation}{0}

In this Section we investigate the behaviour of the energy levels in
string spectrum at large quark masses. To this end, let us
separate the integer part in each root of Eq.~(2.12)
\begin{equation}
M_0\,R\,\omega_n\,=\,(n-1)\pi\,+\,\varepsilon_n,\quad
0\,<\,\varepsilon_n \,<\,\pi, \quad n\,=\,1,2,\dots\,{,}
\end{equation}
where the correction $\varepsilon_n$ to the frequencies $\omega_n$ 
depends on quark masses. By substituting Eq.~(3.1) into Eq.~(2.12), 
one finds, for arbitrary $n$ and in the limit $\mu\to \infty$, that 
$\varepsilon_n$ satisfies the following equation
\begin{equation}
\tan\varepsilon_n\,\simeq\,\frac{2M_0R}{(n-1)\pi\,
+\,\varepsilon_n}\,\mu^{-1}\,{.}
\end{equation}
It means that $\varepsilon_n \,\to \,0$ when $\mu\,\to \,\infty $.
Solution to Eq.~(3.2) for $n=1$ is
\begin{equation}
\varepsilon_1\,\simeq\,\sqrt{\frac{2M_0R}{\mu}}\,{,}
\end{equation}
For $n>1$ we have
\begin{equation}
\varepsilon_n\,\simeq\,\frac{2M_0\,R}{\mu\,(n-1)\,\pi}\,{.}
\end{equation}

Energy levels of the system are calculated by making use of  
Eq.~(2.13). The energy of the first excited state, $E_1$, is
\begin{equation}
E_1\,=\,M_0\omega_1\,=\,\frac{\varepsilon_1}{R}\,
\simeq  M_0\,\sqrt{\frac{2}{M_0\,R\,\mu}}\,{.}
\end{equation}
For fixed string tension $M_0$ and string length $R$, the energy
$E_1$ decreases for raising values of the dimensionless
parameter $\mu$. From physical point of view, this behaviour of $E_1$
implies the instability of the string with heavy quarks at its ends.

Figure 1 shows the dependence of the dimensionless energy of the first
excited state $E_1/M_0\,=\,\omega_1$ on the dimensionless quark mass
$\mu\,=\,m/M_0$ (bold--face curve). This  curve is obtained for the 
string length of order ofthe hadron size\footnote{When
calculating the interquark potential
generated by string
we have to consider arbitrary length of the string $R$.  From 
frequency equation  (2.12) one can easily deduce the following scaling 
property of its roots. Let the functions $f_n(\mu),\,n\,=\,1,2,\ldots $ 
determine the roots of this equation for given value of the mass 
parameter $\mu $ and for $M_0\,R\,=\,1,$ i.e.
$$
\omega_n(\mu,\,M_0R=1)\,=\,f_n(\mu)\,{.}
$$
Then for arbitrary string length $M_0R \,\not= \,1$
the following scaling rule holds
$$
\omega_n(\mu,\,M_0R\not=1)\,=\,\frac{1}{M_0R}\,f_n\left 
(\frac{\mu}{M_0R}\right )\,{.}
$$}
$M_0\,R\,=\,1$. When
$\mu\,\sim\,10^2$ (region of the top quark mass) we have
$E_1/M_0\,\sim\,0.1$.
There are no any restrictions on the number of excitations with 
energy $E_1$. Therefore, we really have infinite sequence of 
equidistant energy levels $E_n\,=\,n\,E_1,\,n\,=1,2,\dots,$ which is 
superimposed on the basic string spectrum with scale $M_0$ 
(see Eq.~(2.13)). As a consequence, for sufficiently large quark masses 
the ground state of the string becomes fuzzy, i.e. weak, compared with 
$M_0$, external perturbations will result in excitation of the
string (see Fig.~1). Obviously this implies the infrared instability 
in this system.

It should be noted that all these considerations are applicable to 
the rigid string too. In fact, in Ref.~\cite{LANE} the same frequency 
equation as (2.11) was derived in the model of the rigid string with 
massive ends.

Where may this infrared instability  be manifested? At first, when 
going beyond
the linear approximation in string model. To this end we have to take 
into account the "interaction" in the string action (2.4) which is 
described by terms $\sim\,{\bf u}^n$ with $n>2$. Besides, this 
infrared instability may be revealed in calculations at finite 
temperature. For example, effective string
potential generated by linearized string action (2.4) is obtained by
functional intagration over "field" ${\bf u}(t,r)$. As a result, one 
arrives at the following well known expression~\cite{KAPU}
\begin{equation}
\sum_{m=-\infty}^{\infty}\sum_{n=1}^{\infty}\,
\ln(\bar\omega_m^2\,+\,k_n^2)\,=\,
\frac{2}{T}\,\sum_{n=1}^{\infty}\left[\frac{k_n}{2}\,+
\,T\,\ln(1\,-\,e^{-k_n/T})\right]\,{,}
\end{equation}
where $\bar\omega_m\,=\,2\pi\,T\,m$ are the Matsubara frequencies and
$k_n/M_0\,=\,\omega_n$ are the
roots of Eq.~(2.12). When $m\to\infty$, $k_1\to 0$ and we obtain
infinity in (3.7). On the other hand, calculations in string models 
with fixed
ends are well defined at finite temperature ~\cite{ANTI}.

\section{Conclusion}
\setcounter{equation}{0}

The instability of the relativistic string, connecting
heavy quarks is quiet clear because fixed string tension $M_0^2$ 
proves to
be insufficient to keep in bound state very heavy quarks. In the 
framework of the string approach to hadron physics this instability 
apparently implies
that string--like collective excitations of gluon field do not 
dominate in heavy quark dynamics. Though other points of view can 
be proposed. For example, one way suppose that
string tension for heavy quarks should be greater than $M_0^2$. 
However the following problem arises here: in what way this 
assumption can be put in agreement with basic concepts of QCD.

In the same way one may doubt in applicability of the linearized 
string action (2.5) which has been used to derive frequency equation 
(2.12). But in this case it would be difficult to understand why 
this action is good for description of the string with light quarks 
at its ends or for fixed string ends  but is
not applicable to heavy quarks. Proceeding from simple physical
considerations one may expect that situation should be opposite 
because linearization of the string action is, in some sense, 
equivalent to the nonrelativistic approximation 
(see, for example, ~\cite{STER}).

Thus the problem of infrared instability in the dynamics of the
relativistic string connecting sufficiently heavy quarks remains, 
in our opinion, open.\\[0.4cm]

{\large \bf Acknowledgements}\\[0.2cm]

One of the authors, V.V. N., would like to thank N.R. Shvetz 
(Yale University)
for valuable discussion of some topics considered in this article. 
G.L. acknowledge the financial support from Bogoliubov Laboratory of
Theoretical Physics (JINR) and the Salerno University. This
work has been partly supported by the Russian Foundation for 
Fundamental Research through Project No. 93-02-3972.

\newpage
\vspace*{4cm}
\centerline{\bf Figure Caption}
\vskip0.8cm

Fig.~1.~Quark mass dependence of the energy levels in string spectrum. The bold--faced
curve is the dimensionless energy of the first excited state, $E_1/M_0$;
$\mu\,=\,m/M_0$ is dimensionless quark mass. Other curves present
the next 5 energy levels, $E_n/M_0\,=\,n\,E_1/M_0,\;\; n\,=\,2,3,\ldots,6.$

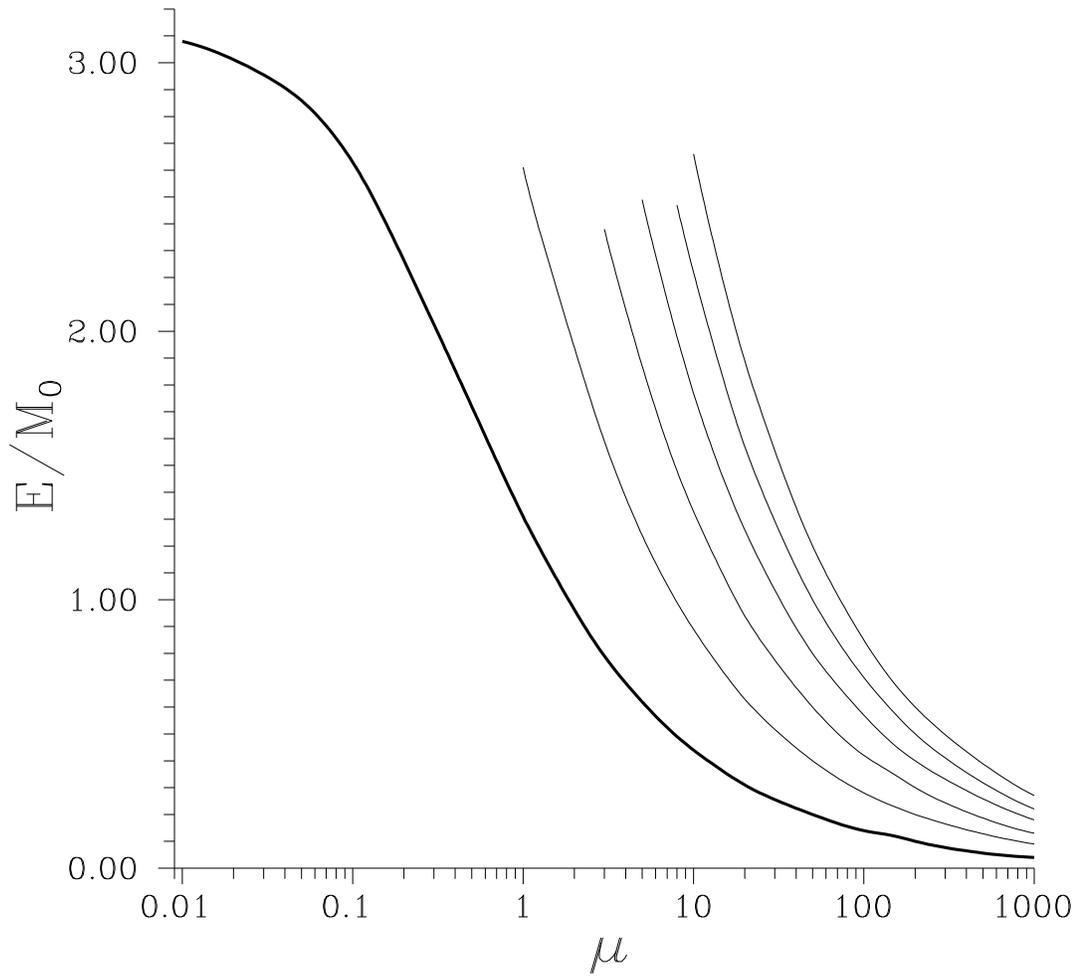
\end{document}